\documentclass[12pt]{article}

\usepackage{amssymb}
\usepackage{amsmath}
\usepackage{graphicx}

\numberwithin{equation}{section}

\begin{document}
\baselineskip=17.5pt
\begin{titlepage}

\begin{center}
\vspace*{10mm}

{\large\bf Power counting renormalizability of quantum gravity
\\
 in Lifshitz spacetime}%
\vspace*{12mm}

Takayuki Hirayama\footnote{taka.hirayama@gmail.com}
\vspace{4mm}

{\it Fukashi High School, Arigasaki 3-8-1, Matsumoto, Nagano 390-8603, 
Japan}\\%
\vspace*{12mm}

\begin{abstract}\noindent%
We analyse the power counting renormalizability of the quantum field theory of Einstein or Einstein-Gauss-Bonnet gravity in $D+2$ dimensional Lifshitz spacetime. We show the spectral dimension becomes $2+(D/z)$ at the UV region where $z$ is the critical exponent. Since it is larger than two, the quantum theory of Einstein gravity is not power counting renormalizable. For the pure Einstein-Gauss-Bonnet gravity, where Lifshitz spacetime is allowed only when the parameters are fine tuned, it happens that the graviton modes do not propagate and the quantum field theory is accidentally renormalizable when $z\geq D$. Another method is discretizing the radial coordinate which changes the spectral dimension to $1+(D/z)$ at the UV region. Since our four dimensional spacetime is continuous, the four dimensional Lorentz symmetry is recovered at the low energy and the power counting renormalizability is still kept for $z\geq D$, if the spacetime near the null singularity in Lifshitz spacetime is modified into AdS spacetime and the discrete radial direction is compactified like a brane world scenario. We also comment on the AdS/CFT correspondence.
\end{abstract}

\end{center}

\end{titlepage}

\newpage

\section{Introduction and Summary}

Einstein gravity is not renormalizable, if Einstein-Hilbert action is quantized, and higher loops induce more and more severe UV divergences in the flat spacetime~\cite{'tHooft:1974bx}. Since the gravitational coupling has the mass dimension $2-n$ in $n$ dimensional spacetime, $L$ loop graphs induce divergences with the power of $\Lambda_{UV}^{(n-2)L+2}$ where $\Lambda_{UV}$ is the UV cutoff in more than two dimensions. For example in four dimensional spacetime ($n=4$), one loop matter graphs induce upto dimension four operators with an appropriate power in the cutoff, i.e. the cosmological constant, Einstein Hilbert term and quadratic terms in Riemann tensor ($R^2$, $R_{ab}R^{ab}$ and $R_{abcd}R^{abcd}$). As long as some symmetries or identities, such as supersymmetry and Bianchi identities, do not forbid such terms, they are induced. Since Einstein gravity does not have quadratic terms in the action, we can not cancel the divergences in front of the quadratic terms in Riemann tensor and the quantum theory is not renormalizable. 

If we start from the higher derivative gravity in four dimensions\footnote{
In four dimensions, $\sqrt{-g} \{ R^2 -4 R_{ab}R^{ab} +R_{abcd}R^{abcd} \}$ becomes total derivatives and then $R_{abcd}R^{abcd}$ can be rewritten by $R^2$ and $R_{ab}R^{ab}$.},
\begin{align}
 S &= \!\! \int \! \! d^4 \! x \sqrt{-g} \left[ M^{n-2} ( R-2\Lambda) +\alpha R^2 +\beta R_{ab}R^{ab} + \gamma R_{abcd}R^{abcd} \right],
\end{align}
the gravitational coupling has the mass dimension zero and we can cancel the divergences using the coefficients in front of quadratic terms in Riemann tensor ($\alpha$, $\beta$ and $\gamma$). Thus the theory is renormalizable~\cite{Stelle:1976gc}, however, the graviton propagator behaves $1/k^2(k^2-m^2)$, where $m$ is computed from the action, and this propagator has additional poles, in addition to the one for the massless spin two graviton, which corresponds to massive spin two graviton and massive spin zero graviton. The massive spin two graviton has a negative kinetic energy, thus a ghost particle, and then the unitarity and/or stability of the flat spacetime is lost\footnote{If we choose the parameters $\alpha$, $\beta$ and $\gamma$ such that the mass of ghost spin two fields becomes infinite, the renormalizability is lost.}. Thus non renormalizability remains as one of important problems in quantum field theory of gravity and this has  lead many thoughts in the concept of spacetime at the short distance.

Horava~\cite{Horava:2009uw} recently proposed an idea that Lorentz symmetry is broken at UV region, but a theory possesses an invariance under the anisotropic rescaling with a dynamical exponent~$z$,
\begin{align}
 t \rightarrow \lambda t , \hspace{3ex}
 x^i \rightarrow \lambda^{1/z} x^i.
\end{align}
A theory with this invariance is a Lifshitz scalar theory~\cite{lifshitz} whose action is
\begin{align}
 S &= \!\! \int \! dt d^D \! x \left[ (\partial_t \phi)^2 - ( \partial_i^{\: z} \phi)^2 \right].
\end{align}
Then the propagator is given $1/(\omega^2 -\vec{k}^{2z})$ where $\omega$ and $\vec{k}$ are the momentum along time and spacial directions respectively. Then the propagator does not contain any additional poles and dumps quickly along large spacial momentum ($z>1$) resulting that the spectral dimension becomes $1+(D/z)$ at the UV region instead of $1+D$~\cite{Horava:2009if}. Therefore the UV divergences are suppressed, and Horava constructed an action of renormalizable quantum gravity with this invariance. 

On the other hand, there is a spacetime, Lifshitz spacetime~\cite{Koroteev, Kachru:2008yh, Taylor:2008tg}, whose isometries match with the anisotropic rescaling,
\begin{align}
 ds^2 &= L^2 \left( -\frac{dt^2}{r^2} + \frac{dx_D^2}{r^{2/z}} + \frac{dr^2}{z^2r^2} \right).
 \label{83}
\end{align}
This metric is invariant under the change $t\rightarrow \lambda t$, $x^i\rightarrow \lambda^{1/z}x^i$ and $r\rightarrow \lambda r$. This metric can be a solution of, for example, Einstein gravity with massive gauge fields or higher derivative gravity~\cite{Kachru:2008yh}. Then we expect that the quantum gravity theory (e.g. $L=\sqrt{-g}(R-2\Lambda) +L_{\rm matter}$) in Lifshitz spacetime shares the same improvement of UV divergences. However this is not clear at all, since the action does not have higher spacial derivatives and the spacetime curvature is naively negligible at the short distance. As analysed in a separate paper~\cite{hirayama} where we analysed a scalar theory in Lifshitz spacetime, the same improvement of UV divergences is achieved after the radion field is integrated out. 

We in this paper study the quantum gravity theories in Lifshitz spacetime, and analyse the propagator. We use Kaluza-Klein picture after compactifying $r$ direction by introducing a cutoff at large $r$, ($r=R$). This is because it is easier to study the UV behaviour of the propagator and we avoid the null singularity appears at $r=\infty$ in the metric~\cite{Kachru:2008yh}. Similar to the scalar theory, we show the propagator at the UV region behaves $1/(\omega^2-\vec{k}^{2z} -k_r^2)$, $k_r$ represents the momentum along $r$ direction, after integrating out the radion fields and the spectral dimension becomes $2+(D/z)$ at UV. Thus the UV behaviour is improved, but since the spectral dimension is still larger than two, the theories are still power counting non-renormalizable. However it happens that in the pure Einstein-Gauss-Bonnet gravity theory the graviton does not propagate (does not have time derivative terms in the kinetic terms) and the theory is accidentally power counting renormalizable when $z\geq D$. This is because Lifshitz spacetime is a solution of Einstein equation only if the parameters, the cosmological constant and the coefficient in front of Gauss-Bonnet term, are fine tuned~\cite{Adams:2008zk}. One may try to think whether supersymmetry or higher derivative gravity will help the situation and reduce the spectral dimension. Supersymmetry forbids some terms, but allows some higher dimensional terms which are induced with divergent coefficients in the end from higher loops as long as the spectral dimension is larger than two. If we limit ourselves to Lovelock theories in order to avoid higher time derivatives in the linearized Einstein equation, the behaviour along $r$ direction does not change and the theory is not power counting renormalizable. 

However if we discretize $r$ direction and introduce the cutoff for the largest $k_r$, we can reduce the spectral dimension by one. Although it has been known that cutting infinite tower of massive Kaluza Klein graviton into finite number of massive Kaluza Klein graviton is inconsistent, recent development~\cite{Hinterbichler:2011tt} in building consistent theories of massive spin two  fields coupled to gravity makes us believe consistent gravity theories of finite number of massive spin two fields exist. Therefore we assume in this paper that discretization of $r$ direction is possible and discuss what can be expected. Then we can discuss that although the isometry under the anisotropic rescaling is broken,  a discrete one remains and the UV behaviour $1/\vec{k}^{2z}$ does not change. 
Then the spectral dimension becomes $1+(D/z)$ and the quantum theory becomes renormalizable when $z\geq D$. Although the discretization explicitly breaks the $D+2$ dimensional diffeomorphism, this keeps the $D+1$ dimensional one and the $D+1$ dimensional Lorentz symmetry is recovered at the low energy when the spacetime flows from Lifshitz spacetime into AdS spacetime toward $r=\infty$ and compactify the discrete radial direction as we will see. In the $D+1$ dimensional effective theory, this corresponds to introducing lower spacial derivatives $(\partial_i^{\: n}\phi)^2$, $n=1,\cdots,z-1$, ($\phi$ represent a general field) which is consistent with AdS/CFT correspondence~\cite{Maldacena:1997re}. As long as the spacetime is not exactly AdS space, but is Lifshitz spacetime at small $r$ and approaches to AdS spacetime at large $r$, the UV behaviour does not change. Thus applying this theory for $D=3$ and $z=3$, we construct a renormalizable brane world scenario in which SM particles are placed in the bulk or at a brane.


\section{Quantum field theory of Einstein gravity in Lifshitz spacetime}

We study the graviton propagator in Lifshitz spacetime in order to analyse the degree of UV divergences for a general Feynman graph in quantum field theory of Einstein gravity. Lifshitz spacetime can be a solution in, for example, Einstein gravity with massive gauge fields or higher derivative gravity theories. Here we use Einstein gravity with massive gauge fields in $D+2$ dimensions. The action is
\begin{align}
 S &= \!\! \int \!\! dt d^D \! x dr \sqrt{-g} \left[ M^D ( R-2\Lambda) - \frac{1}{4}F_{ab}F^{ab} 
 -\frac{m^2}{2}A_a A^a \right], 
 \label{103}
\end{align}
where we ignored the surface terms. If one wants, one can embed this theory into an Abelian Higgs model. Einstein equation is 
\begin{align}
 M^D\left\{ R_{ab} - \frac{1}{2} g_{ab} (R-2\Lambda) \right\} &= \frac{1}{2}F_{ac}F_{b}^{~c} +\frac{m^2}{2}A_aA_b
 -\frac{1}{2}g_{ab} \left\{ \frac{1}{4} F_{cd}F^{cd} +\frac{m^2}{2}A_cA^c \right\}
 .
\end{align}
The following Lifshitz metric and the gauge field configuration solve Einstein equation and the equation of motion for the gauge fields,
\begin{align}
 ds^2 &= L^2 \left( -\frac{dt^2}{r^2} + \frac{ dx_D^2}{r^{2/z}} + \frac{dr^2}{z^2r^2} \right) , 
 \hspace{3ex}
 A_t(r) = \frac{q}{r},
 \label{115}
\end{align}
where
\begin{align}
 \Lambda = -\frac{1}{2L^2} ( D^2 +(D-1)z +z^2) , \hspace{3ex}
 m^2 = \frac{Dz}{L^2}, \hspace{3ex}
 q = \sqrt{\frac{2(z-1)}{z}} L M^{D/2}.
\end{align}
This metric has an isometry under the rescaling,
\begin{align}
 t \rightarrow \lambda t, \hspace{3ex}
 x^i \rightarrow \lambda^{1/z} x^i, \hspace{3ex}
 r \rightarrow \lambda r,
 \label{128}
\end{align}
and the gauge field $A_a dx^a$  is invariant.
Then the momentum of a plane wave $e^{i(\omega t -k_i x^i -k_r r)}$ has the corresponding rescaling, $\omega \rightarrow \lambda^{-1} \omega$, $k_i \rightarrow \lambda^{-1/z} k_i$ and $k_r \rightarrow \lambda^{-1}k_r$ and the propagator is then expected to have a form of $1/(\omega^2-(\sum k_i^2)^z-k_r^2)$. Then this propagator dumps as quick as $1/(\sum k_i^2)^z$ along a large spacial momentum and UV divergences are suppressed for $z>1$. However since the action does not contain higher spacial derivatives, we are not sure whether this symmetry argument works.
Therefore we expand the metric and gauge fields around Lifshitz background and study the linearlized Einstein equation and equation of motion for the gauge fluctuations in order for analysing the propagator of graviton and gauge fields. 

We expand the metric and gauge fields,
\begin{align}
 g_{ab} &= {\rm diag} \left( -\frac{L^2}{r^2} , \frac{L^2}{r^{2/z}}, \cdots, \frac{L^2}{z^2r^2} \right) +\widetilde{h}_{ab},
 \label{138}
 \\
 A_{a} &= \frac{q}{r} \delta_a^t +\delta A_a.
\end{align}
In computation, we take Gaussian normal gauge, i.e. $\widetilde{h}_{ar}=0, (a=t,x^1,\cdots,x^D,r)$. Since the diffeomorphism $\delta \widetilde{h}_{ab} = \nabla_a \widetilde{\xi}_b +\nabla_b \widetilde{\xi}_a$ becomes
\begin{align}
 \delta \widetilde{h}_{tr} &= \frac{1}{2r}\left\{ (r\partial_r +2)\widetilde{\xi}_t + r \partial_t \widetilde{\xi}_r \right\} ,
 \\
 \delta \widetilde{h}_{ir} &= \frac{1}{2zr} \left\{ ( zr\partial_r +2 )\widetilde{\xi}_i -z r \partial_i \widetilde{\xi}_r \right\},
 \hspace{3ex} i = 1,\cdots, D,
 \\
 \delta \widetilde{h}_{rr} &= \frac{1}{2r} (r\partial_r +1 ) \widetilde{\xi}_r ,
\end{align}
Gaussian normal gauge is achieved by using the gauge parameters, $\widetilde{\xi}_t,\widetilde{\xi}_1,\cdots,\widetilde{\xi}_D$ and $\widetilde{\xi}_r$. Although we still have remaining gauge freedom, we do not fix the fluctuations more. 

Hereafter we take $D=3$ and study a plane wave solution propagating along $x^1$ direction. Thus we take $\widetilde{h}_{ab}= \widetilde{h}_{ab}(r) e^{i(\omega t - kx^1)}$ and $\delta A_a = \delta A_a(r) e^{i(\omega t -kx^1)}$ (we should take the real parts in the end). Moreover we write
\begin{align}
  \tilde{h}_{ab}
 &=
 L^2\left( \begin{array}{ccccc}
  \displaystyle \frac{h_{tt}(r)}{r^2} & \displaystyle\frac{h_{t1}(r)}{r^{1+(1/z)}} & \displaystyle\frac{h_{t2}(r)}{r^{1+(1/z)}} 
  & \displaystyle\frac{h_{t3}(r)}{r^{1+(1/z)}} & 0\\
  \displaystyle\frac{h_{1t}(r)}{r^{1+(1/z)}} & \displaystyle\frac{h_{11}(r)}{r^2} & \displaystyle\frac{h_{12}(r)}{r^{2/z}} 
  & \displaystyle\frac{h_{13}(r)}{r^{2/z}} & 0 \\
  \displaystyle\frac{h_{2t}(r)}{r^{1+(1/z)}} & \displaystyle\frac{h_{21}(r)}{r^{2/z}} & \displaystyle\frac{h_{22}(r)}{r^2} 
  &\displaystyle \frac{h_{23}(r)}{r^{2/z}} & 0 \\
  \displaystyle\frac{h_{3t}(r)}{r^{1+(1/z)}} & \displaystyle\frac{h_{31}(r)}{r^{2/z}} & \displaystyle\frac{h_{32}(r)}{r^{2/z}} 
  & \displaystyle\frac{h_{33}(r)}{r^2} & 0 \\
  0 & 0 & 0 & 0 & 0
\end{array}
  \right) e^{i(\omega t -kx^1)},
  \label{169}
  \\
  \delta A_a &= 
  \left( \frac{a_t(r)}{r}, \frac{a_{1}(r)}{r^{1/z}}, \frac{a_{2}(r)}{r^{1/z}},\frac{a_{3}(r)}{r^{1/z}},
  \frac{a_r(r)}{r} \right) e^{i(\omega t -kx^1)} ,
\end{align}
so that $h_{ab}(r)$ and $a_a(r)$ are invariant under the anisotropic rescaling.

The  equations are complicated and lengthy. The independent modes are divided into vector and scalar modes under the remaining $SO(2)$ symmetry which rotates $(x^2,x^3)$ directions.
The nonzero components of vector modes are
\begin{align}
 h_{t2}(r) &= \omega r v_1(r) + p r^{1/z} v_2(r), \hspace{3ex}
 h_{12}(r) &= -p r^{1/z} v_1(r) +\omega r v_2(r), \hspace{3ex}
 a_2(r),
 \\
 h_{t3}(r) &= \omega r v_3(r) + p r^{1/z} v_4(r), \hspace{3ex}
 h_{13}(r) &= -p r^{1/z} v_3(r) +\omega r v_4(r), \hspace{3ex}
 a_3(r).
\end{align}
$v_2(r)$ and $a_2(r)$ satisfy two coupled second $r$ derivative differential equations and thus there are two independent modes which are spin two graviton and spin one mode.  $v_1(r)$ satisfies a first differential equation coupled with $v_2(r)$ and $a_2(r)$, but using a remaining gauge freedom $v_2(r)$ does not give an independent mode. If we replace $(v_1(r),v_2(r),a_2(r))$ by $(v_3(r),v_4(r),a_3(r))$ in the equations, the equations become those for $(v_3(r),v_4(r),a_3(r))$. Thus there are spin two graviton and spin one mode in $v_4(r)$ and $a_3(r)$. 

One of scalar modes is given by $h_{23}(r)$ which satisfies
\begin{align}
 z^2 r^2 h''_{23}(r) - 3zr h'_{23}(r) + (\omega^2 r^2 - k^2 r^{2/z} ) h_{23}(r) &=0,
 \label{181}
\end{align}
and this is a spin two graviton mode. We find four more scalar modes, and  two of them are spin two graviton modes and the other two are the spin one modes.

In total we find five spin two modes and four spin one modes as expected. We then solve these equations numerically. However, the equations are qualitatively same as the equation for a scalar field in Lifshitz spacetime. The free scalar theory in Lifshitz spacetime~is
\begin{align}
 S &= \!\! \int \!\! dt d^D \! x dr \sqrt{-g} \left[ -(\partial \phi)^2 -m^2 \phi^2 \right].
\end{align} 
We substitute Lifshitz metric~\eqref{115} into the action and compute the equation of motion;
\begin{align}
 \left[ -\partial_{\widetilde{r}}^2 +V(\widetilde{r}) \right] \widetilde{\phi}_{(\omega,\vec{k})}(\widetilde{r})
 &= E \widetilde{\phi}_{(\omega,\vec{k})}(\widetilde{r}),
 \label{143}
 \\
 V(\widetilde{r}) &= \frac{D^2+2Dz+4L^2m^2}{4z^2} \frac{1}{\widetilde{r}^2} +\frac{1}{z^2} \frac{1}{\widetilde{r}^{2-(2/z)}},
 \\
 E&= \omega^2 k^{-2z} z^{-2} ,
\end{align}
where we fourier expand the scalar field as  
\begin{align}
 \phi(t,x^i,r) &= \!\! \int \!\! d\omega d^D\! k \: \widetilde{r}^{D/2z}\widetilde{\phi}_{(\omega,\vec{k})}(\widetilde{r})e^{i(\omega t +k_i x^i)} ,
\end{align}
and $\widetilde{r}=k^z r$ and $k^2=\sum_i k_i^2\neq 0$. In~\eqref{181} we define $h_{23}(r)= r^{3/(2z)} \phi(r)$ and obtain
\begin{align}
 -\phi''(r) + V(r)\phi(r) &= E\phi(r),
 \\
 V(r) &= \frac{9+6z}{4z^2}\frac{1}{r^2}+ \frac{1}{z^2}\frac{k^2}{r^{2-(2/z)}},
 \\
 E &= \omega^2 z^{-2},
\end{align}
and thus this is same as~\eqref{143} with $D=3$ and $m=0$ after defining $\widetilde{r}=k^zr$.

The other modes satisfy complicated equations of motion, but if we take large $r$, $\omega k$ and $k^2$ terms drop out and the equations become simplified. We find that at large $r$ region all spin two modes satisfy the same equation with $m=0$ in~\eqref{143}, and all spin one modes satisfy~\eqref{143} with $m=\sqrt{2(z^2-1)}/L$. Since the KK spectrum is mainly determined from the potential at the large $r$, the Kaluza Klein spectrum for spin two and zero modes is qualitatively same as that for the scalar field.

As discussed in the separate paper, it is easy to study the UV behaviour of the propagator from the Kaluza-Klein picture by compactifying $r$ direction. We then introduce the cut at large $r$, $r=R$, and numerically computed the KK spectrum for the scalar field with $D=3$, $z=4$ and $m=0$. The results of KK masses are given as
\begin{align}
 \omega_n^2 &= \frac{c_n}{R^{2-(2/z)}} k^2 + \frac{d_n}{R^2},
\end{align}
and $c_n$ and $d_n$ behave $n^{2-(2/z)}$ and $n^2$ for large $n$. Then the effective $D+1$ dimensional action for the kinetic terms become
\begin{align}
 S_{D+1} &= \!\! \int \!\! dt d^D \! x \sum_n \left[ (\partial_t \phi^{(n)} )^2 - \frac{c_n}{R^{2-(2/z)}} (\partial_i \phi^{(n)})^2
 -\frac{d_n}{R^2} ( \phi^{(n)} )^2 \right],
\end{align}
where we denote $\phi^{(n)}$ as spin two or one modes. Then this gives a propagator of $1/(\omega^2 -k^2 -m^2)$ and the propagator does not show $1/k^{2z}$ behaviour at large spacial momentum. However here we should take into account the effects of the radion fields. Since the metric has the isometry under the anisotropic rescaling~\eqref{128}, there is a massless mode associated with the isometry, the radion, which is a scalar mode in the spin two mode and causes the shift of $R$. Then the equation of motion for the radion fields are read from that for $R$ by treating $R$ as a field. Then the equation is schematically written as
\begin{align}
 R^{2/z} \sim \left( \sum_m d_m ( \phi^{(m)} )^2 \right) \left/
  \left( \sum_n c_n ( \partial_i \phi^{(n)} )^2 \right) \right. \sim N^{2/z} / (\partial_i)^2
\end{align}
where we introduced the cut $N$ for the number of eigenmodes and replaced $(\phi^{(n)})^2$ with the VEV. Then plugging this into the action, we have
\begin{align}
 S_{D+1} \sim \!\! \int \!\! d^{D+1}  x \sum_n \left[ (\partial_t \phi^{(n)} )^2 - \widetilde{c}_n (\partial_i^{\: z} \phi^{(n)} )^2 \right],
\end{align}
where $\widetilde{c}_n \sim n^2/N^2$, ( or $\sim n^{2-(2/z)}/N^{2-(2/z)}$) and thus $\widetilde{c}_n$ is small for small $n$ and becomes order one for large $n$. Although this computation is rough, the two point function after summing up the Feynman graphs whose intermediate state contains the radion fields becomes $1/(\omega^2-k^{2z})$. In total, the propagator of graviton and gauge fields in $D+2$ dimensions behaves
\begin{align}
 \frac{1}{\omega^2-k^{2z} -k_r^2}
\end{align}
at UV after taking into account the contribution of radion fields and $k_r$ represents the momentum along $r$ direction ($n^2/R^2\rightarrow k_r^2$). Then one can show the spectral dimension is $2+(D/z)$ since if we have a loop integral, we define $s_i=k_i^z$ and
\begin{align}
 \int \!\! d\omega d^D \! k dk_r  \frac{1}{\omega^2-k^{2z} -k_r^2}
 = \!\! \int \! \! d\omega dk_r \left( \prod_{i=1}^{D} ds_i s_i^{-1+(1/z)} \right) \frac{1}{\omega^2-s^2-k_r^2} .
 \label{251}
\end{align}
Now we compute the degree of UV divergences $\Gamma$ for a Feynman graph for the theory~\eqref{103},
\begin{align}
 \Gamma &\leq \left\{ D+2 - D(1-z^{-1} ) \right\}L -2I +2V
 = (D/z)L +2,
 \label{259}
\end{align}
where we use the relation $I=V+L-1$ and $L$, $I$ and $V$ are the number of loops, internal lines and vertex. The gravitational interactions have up to second derivatives and we have $2$ in front of $V$ in the computation and we have $-2$ in front of $I$ since we use $s_i$ in the propagator~\eqref{251}. Since the spectral dimension is $2+(D/z)$, the quantum field theory of Einstein gravity in Lifshitz spacetime is power counting non-renormalizable. For example, $L\geq 2z/D$ loop graphs induce $R^2$, $R_{ab}^2$ and $R_{abcd}R^{abcd}$ with divergent coefficients and we cannot renormalize these divergences. 

We can see from the expression~\eqref{259} that the quantum gravity theory is renormalizable if the spectral dimension is less than or equal to two. Then if we take $z=D$, the spectral dimension becomes three and we come to think that the theory becomes renormalizable if we can modify the theory along $r$ direction so that the spectral dimension is reduced by one. In the following section, we study other theories.


\section{Quantum field theory of Einstein-Gauss-Bonnet gravity in Lifshitz spacetime}

We studied the spectral dimension becomes $2+(D/z)$ at UV, and wonder whether Lovelock gravity changes the behaviour of the propagator. In this section we study Einstein-Gauss-Bonnet theory. 

Lifshitz spacetime is a solution when the parameters are fine tuned in Einstein-Gauss-Bonnet theory,
\begin{align}
 S &= \!\! \int \!\! dt d^D \! x dr \sqrt{-g} \left[ M^D (R-2\Lambda ) +a (R^2-4R_{ab}R^{ab} +R_{abcd}R^{abcd}) \right].
\end{align}
Einstein equation, 
\begin{align}
 M^D(R_{ab}-\frac{1}{2}g_{ab}(R-2\Lambda) )
 +a \Big[
 -\frac{1}{2}g_{ab}   (R^2 -4R_{cd}R^{cd} + R_{cdef}R^{cdef})
 &\nonumber\\
 +2RR_{ab} -4 R_{ac}R_b^{~c}
 -4R_{acbd}R^{cd}
 +2R_{acde}R_{b}^{~cde} \Big]
 &=0,
\end{align}
is solved with Lifshitz metric if the coefficient $a$ is fine tuned as
\begin{align}
 ds^2 &= L^2 \left( -\frac{dt^2}{r^2} + \frac{dx_D^2}{r^{2/z}} + \frac{ dr^2}{r^2} \right)
 , \hspace{3ex}
 \Lambda = -\frac{(D+1)D}{L^2} , \hspace{3ex}
 a = \frac{L^2M^D}{2(D-1)(D-2)},
\end{align}
where we have set $z\neq 1$.
We then expand the metric, as is done in~\eqref{138} and~\eqref{169}, and compute linearlized Einstein equation for analysing the propagator. Then in the case of $D=3$, we find that $h_{tt}(r)$, $h_{t1}(r)$, $h_{t2}(r)$, $h_{t3}(r)$ vanish from linearized Einstein equation, $h_{12}(r)=h_{13}(r)=0$ is the solution (we used the remaining gauge freedom.), and 
\begin{align}
 h''_{23}(r) &= \frac{3}{zr} h'_{23}(r) + \frac{z+1}{z^2r^2}k^2r^{2/z}h_{23}(r),
 \\
 h'_{33}(r) &= - h'_{22}(r),
 \\
 h'_{11}(r) &= -\frac{1}{zr} k^2 r^{2/z} ( h_{22}(r) + h_{33}(r) ),
 \\
 z^2r^2 h'''_{22}(r) &+ zr(2z-5)h''_{22}(r) + (3(2-z) - k^2 r^{2/z}(z+1) ) h'_{22}(r) =0.
\end{align}
Thus the time derivatives vanish (there is no $\omega^2$ term) and there are no propagating modes in time. Therefore this theory does not have interesting dynamics although the spectral dimension is reduced by one (time direction disappears) and the theory is then renormalizable. This happens since Lifshitz spacetime is allowed only when the parameters are fine tuned. Thus if the metric is slightly modified from Lifshitz metric, the time derivative terms are immediately induced.

\vspace{3ex}

Lifshitz spacetime is still a solution when we add Gauss-Bonnet term in Einstein gravity with massive gauge fields~\eqref{103}. In this case ($D=3$), the equation of motion for $h_{23}(r)$, for example, becomes
\begin{align}
-\phi''(r)+V(r)\phi(r) &= E\phi(r),
\\
V(r)&= \frac{9+6z}{4z^2} \frac{1}{r^2}+\frac{L^2M^3-4az^2}{L^2M^3-4az}\frac{1}{z^2} \frac{k^2}{r^{2-(2/z)}},
\\
E&= \frac{L^2M^3-4a}{L^2M^3-4az}\frac{\omega^2}{z^2},
\end{align}
where $h_{23}(r)=r^{3/(2z)}\phi(r)$, and $a$ is the coefficient in front of the Gauss Bonnet term. Then the Gauss-Bonnet term, as expected, does not change the qualitative behaviour of the propagator at UV.

One may expect that there is a combination of quadratic curvature terms which induces higher $r$ derivative, but does not induce higher time derivatives. However one can check that this is not the case. It is interesting to search more general higher curvature terms for finding a combination which induces higher $r$ derivatives and does not induce higher time derivatives. If this is possible, the UV behaviour of graviton propagator changes along large $k_r$ region and the theory is expected to be renormalizable.


\section{Discrete radial coordinate}

A simple idea of suppressing the propagator along large $r$ momentum is introducing an UV cutoff for largest  $r$ momentum. Although it has been known that truncating infinite Kaluza-Klein graviton modes into finite number of Kaluza-Klein massive graviton is inconsistent, recent development in building consistent theories of massive spin two fields couple to gravity~\cite{Hinterbichler:2011tt} makes us believe there exists consistent theories of finite number of massive spin two fields in a curved spacetime. Therefore in this paper we assume it is possible to discretize $r$ coordinate, and we discuss what is implied. 

Since we discretize $r$ direction, the isometry under the anisotropic rescaling~\eqref{128} is broken. However we can introduce the lattice in such a way that the lattice spacetime still has a symmetry under the discrete anisotropic rescaling. Then we expect that the radion fields exist and the modification of propagator is suppressed by the size of the discretization. Therefore the UV behaviour of propagator does not change, and the spectral dimension now becomes $1+(D/z)$ since discretization reduces the spectral dimension by one. Then this becomes two or less than two when we take $z\geq D$, and the quantum gravity theory becomes renormalizable. We do not discretize $D+1$ dimensions, in which our four dimensional spacetime exists, and then we recover the $D+1$ dimensional Lorentz symmetry by flowing the spacetime from Lifshitz spacetime into AdS$_{D+2}$ spacetime toward $r=\infty$. In terms of the metric in~\eqref{83}, $g^{ii}=L^{-2}r^{2/z}$ changes into $g^{ii}=L^{-2} ( r^{2/z}+b_{z-1}r^{2/(z-1)}+\cdots + b_1r^2)$. At small $r$, $r^{2/z}$ is dominant in $g^{ii}$ and the spacetime is approaching to Lifshitz spacetime and at large $r$, $r^2$ is dominant and the spacetime is approaching to AdS$_{D+2}$ spacetime. Then the Kaluza-Klein modes satisfy $\omega_n^2= c^{(z)}_n k^2/R^{2-(2/z)} +c_n^{(z-1)} k^2/R^{2-2/(z-1)}+\cdots +c_n^{(1)}k^2+ d_n/R^2$. Then repeating the same argument, i.e. taking into account the contribution from the radion field by solving the equation of motion for $R$, we have lower spacial derivative terms in the action $(\partial_i^{\: n} \phi)^2$ with $n=1,2,\cdots, z-1$. Therefore at low energy the $D+1$ dimensional Lorentz symmetry is recovered. This is consistent with AdS/CFT correspondence in which large $r$ corresponds to IR region in CFT side. As the radial coordinate becomes larger, larger power in $r$ becomes important in $g^{ii}$ which induces more lower spacial derivatives in the $D+1$ dimensional effective action until $(\partial_i\phi)^2$ is induced. Since even at large $r$, $r^{2/z}$ component in $g^{ii}$ remains and so the UV behaviour of the propagator does not change anywhere in the bulk and the spectral dimension remains $1+(D/z)$. Therefore the renormalizability is not lost by this modification. In addition, the flow of spacetime has another merit which is that this flow resolves the null singularity appears at $r=\infty$ in Lifshitz spacetime. 

We take $D=3$ and $z\geq 3$, and construct a renormalizable brane world scenario by compactifying the discrete radial direction so that the weak scale is dynamically generated by the warp factor and by introducing Standard Model particles in the bulk or at a brane.


\section{Discussion}

We studied the propagator for graviton in Lifshitz spacetime and show it becomes $1/k^{2z}$ at large spacial momentum and this is true for scalar and gauge fields as well. Although we naively expect that at a short distance the curvature of spacetime is negligible and the UV behaviour of the propagator does not change from $1/k^2$, it does change after taking into account the effects of the radion fields by solving the equation of motion for the radion fields. Since this computation is very rough, it is worth doing a lattice calculation, for example, to confirm this behaviour. 

The spectral dimension becomes $2+(D/z)$ at UV in which $2$ comes from time and radial directions and $D/z$ comes from $D$ dimensional spacial directions. This is larger than $2$ and the quantum gravity is not renormalizable. If one has higher radial derivatives in the action, one can reduce the spectral dimension and the quantum gravity theory becomes renormalizable. However Gauss-Bonnet gravity theory and other Lovelock theories do not induce higher radial derivatives. It is interesting to search a combination of higher curvature terms in Riemann tensor which induces higher radial derivatives, but does not induce higher time derivatives in Lifshitz spacetime.

We discretize the radial direction in order to reduce the spectral dimension. The existence of radion field is important for improving the UV behaviour of the propagator and we do not need to recover the full $D+2$ dimensional diffeomorphism in the continuous limit. Therefore it may be relatively easy to construct a consistent gravity theory in the discrete radial space.

It is also interesting to look for other curved spacetimes in which the spectral dimension at UV become two or smaller than two.


\subsection*{Acknowledgments}
\noindent
I thank my wife for the encouragement.

\end{document}